\documentclass[epj]{svjour}
\usepackage{epsfig}
\begin{document}
\title{ac driven sine-Gordon solitons: dynamics and stability}
\author{Niurka R.\ Quintero \and Angel S\'anchez}
\institute{Grupo Interdisciplinar de Sistemas Complicados,
Departamento de Matem\'aticas, Universidad Carlos III de Madrid,\\
E-28911 Legan\'{e}s, Madrid, Spain.
\email{kinter@math.uc3m.es,\,anxo@math.uc3m.es}}
\date{Received: \today / Revised version: }
\abstract{
The ac driven sine-Gordon equation is studied analytically and numerically,
with the aim of providing a full description of how soliton solutions behave. 
To date, there is much controversy about when ac driven dc 
motion is possible. Our work shows that kink solitons 
exhibit dc or oscillatory motion depending on the relation between their 
initial velocity and the force parameters. Such motion is proven to be 
impossible in the presence of damping terms. 
For breathers, the force amplitude range
for which they exist when dissipation is absent 
is found. All the analytical results are
compared with numerical simulations, which in addition exhibit no dc
motion at all for breathers, and an excellent agreement is found. 
In the conclusion, the generality of our results and connections to
others systems for which a similar phenomenology may arise are discussed.
\PACS{
{03.40.Kf}{Waves and wave propagation: general mathematical aspects} \and
{74.50.+r}{Proximity effects, weak links, tunneling phenomena, and Josephson effects} \and
{85.25.Cp}{Josephson devices}
} 
} 

\maketitle

\section{Introduction}
\label{intro}
In the past two decades, solitons have become a very generic and useful 
paradigm of intrinsically nonlinear phenomena in many different fields 
of physics \cite{general,yo}. Generally 
speaking, solitons can be classified 
\cite{km,sv} in three main 
groups: dynamical, envelope, and topological solitons, according to their
most relevant characteristics such as stability, defining parameters, 
response to perturbations and so on. 
A canonical example of the most important of those groups \cite{nota}, 
topological and envelope solitons, in physical 
systems is the sine-Gordon (sG) equation, widely used 
both because it is integrable 
(meaning that the corresponding initial value problem can always be 
solved analytically in closed form) and because it accurately models 
a variety of problems and applications. With respect to the
integrability of the sG equation, this property
gives us the possibility to know exactly its multisoliton
solutions, that can be classified as kinks (topological) and breathers
(envelope). On the side of the applicability, the sG
equation describes propagation of ultra-short optical pulses in resonant 
laser media \cite{lamb}, a unitary theory of elementary particles 
\cite{skyrme1,skyrme2,enz,raj}, propagation of magnetic flux in Josephson 
junctions \cite{Barone}, transmission of ferromagnetic waves \cite{feld}, 
epitaxial growth of thin films \cite{CW,us}, motion of dislocations in 
crystals \cite{frenkel,Nabarro}, flux-line unlocking in type II
superconductors \cite{nuevo}, DNA dynamics \cite{eng,DNA,yaku}, and many 
others. 

When using the sG equation as a model for an actual physical situation,
it is often necessary to account for factors
that cause deviation from the perfect system, arising, for example, from
forces acting on it, thermal effects, fluctuations, dissipation, or 
spatial modulations (deterministic or random). To account for some or all 
of those, appropriately chosen perturbing terms have to be included 
in the sG equation \cite{km,sv}; to quote a few instances of such 
perturbations, (see also the next paragraph) let us mention 
the studies of forces acting over a DNA molecule, or long 
DNA fragments containing regions of finite size and specific structure
\cite{yaku2}, additive and multiplicative noise sources \cite{pv},
spatially periodic parametric potential \cite{asa} or 
damping with spatiotemporal periodic driving \cite{dca}. 
The work reported on here belongs to this class of problems; specifically,
it focuses on the action of ac (sinusoidal in time, homogeneous in space)
forces on the solitonic solutions, kinks and breathers, of the sG equation.
In mathematical terms, this means we have to deal with the equation
\begin{equation}
\phi_{tt} - \phi_{xx} + \sin(\phi) =
-\beta \phi_{t} + f(t),
\label{ecua1}
\end{equation}
where subindices indicate derivatives with respect to the corresponding 
variables, $\beta\phi_t$ is the usual damping term and 
$f(t)=\epsilon
\sin(\delta t + \delta_{0})$ is an external periodic force describing,
for example, a long Josephson junction under the application of a uniform 
microwave field \cite{Barone}. From now on, the amplitude of the force, 
$\epsilon$, will be our perturbation parameter, i.e., it is assumed that
the force is not too large. We will see, however, that we can deal with 
reasonable and useful values of $\epsilon$, so our calculations and results
are not purely academic or mathematical but rather they have physical 
relevance.

The studies of forced sG systems date back to the late seventies, to the 
seminal works of Fogel {\em et al.}\ \cite{fogel} and of McLaughlin and 
Scott \cite{McL} where the ideas of collective coordinate techniques 
(see \cite{yo} for a review on those) were first introduced to deal with
force terms added to the sG equation. Their results, confirmed by numerical
simulations of the full partial differential equation, showed that dc 
(constant both in space and time) forces act on the sG solitons as if 
they were point-like particles, accelerating them up to the maximum velocity
allowed by the equation in the absence of damping or to a terminal velocity 
in the presence of damping. The problem of dc forces was given further 
consideration in subsequent works, such as \cite{lomol,ariyasu}, where 
aspects such as breather instabilities and pattern formation were the 
main subjects; as our principal goal is the study of ac forces, we will 
not go into more detail on those here. 
Dissipation effects on breathers
were first studied in \cite{mc}, where McLaughlin and Overman 
showed that free (i.e, not driven) damped breathers where anihilated, 
their energy being 
dissipated into radiation. Subsequently, Lomdahl and Samuelsen
\cite{lomsam1,lomsam2} showed that the introduction of ac forces in
the damped sG equation compensated those losses and stabilized
breathers, whose frequency became modulated by the driving. This is the 
main result about ac breather dynamics, and if fact we are
not aware of later relevant work on ac driven breather dynamics. 
On the other hand, the kink case
has a longer and more problematic story, that begins with the research by 
Olsen and Samuelsen in \cite{olsam},
where the authors studied the effect of dc and ac forces 
over sG kinks. In their analytical calculation, they considered two special 
cases for an ac force given by $\epsilon \sin(\delta t + \delta_{0})$, namely
$\delta_{0}=0,\pi/2$ with $\delta \ne 1$ and sufficiently small kink
velocities. Their results were obtained in the approach that the solution of 
the perturbed sG equation can be divided into a kink part and a vacuum part. 
Although most of the paper deals with dc forces, the authors conclude that 
for the two cases studied for ac forces, one leads to kink dc motion 
($\delta_{0}=0$) whereas for the other ($\delta_{0}=\pi/2$) only 
oscillatory motion is found. For both choices, the work is further 
restricted to null initial velocity only, $u(0)=0$, with the subsequent 
loss of generality of their findings.
Numerical simulations confirmed the predictions
and the accuracy of the analytical calculations. 
However, no other initial phase differences were studied, and therefore a 
general analysis (including also initial velocity) 
was lacking. In addition, 
they did not present any results for ac driving plus damping, saying simply 
that similar results could be obtained in that case. A few years after that
work, Bonilla and Malomed \cite{boni} 
revisited the same problem and concluded that dc
motion of ac driving kinks was only possible in discrete sG equations (i.e.,
the spatial variable is discretized), and only in the presence of damping. 
Such a result, in contradiction with that in \cite{olsam}, was later shown 
to originate in an incorrect analytical approximation to the problem in 
\cite{cai}; the perturbative theory worked out in that paper suggested 
that kink dc motion induced by ac forces in the presence of damping was 
not possible, and the authors stated their belief hat such phenomenon was 
unlikely in general, i.e., in the undamped case (at that time, 
they were not aware of the results of Olsen and Samuelsen \cite{olsam}).
Finally, the last contribution to this topic prior to our work appeared 
in \cite{zaragoza}, where dc motion of undamped ac driven kinks was 
found in simulations of the Frenkel-Kontorova (discrete sine-Gordon) model.
In this case, the phenomenon was explained in terms of a resonance 
between the external force and internal oscillatory modes of the discrete
kink (absent in the continuum models).

The brief summary in the above paragraph is enough to understand that the
knowledge on the ac driven dynamics was not complete and systematic and,
what is worse, it severely suffered from the contradictory results obtained
about the possibility of dc propagation of kinks. In view of this, we first
addressed this issue in \cite{anniu}, a preliminary work on small 
(non-relativistic, $|u(t)| \ll 1$) velocity, ac driven kinks. Using the 
techniques developed in \cite{McL}, we considered the general ac force 
$\epsilon \sin(\delta t + \delta_{0})$ with and without dissipation. By 
this means, we were able to show analytically that there indeed exists
dc motion of kinks provided dissipation is absent. Moreover, we showed that
the velocity of such dc motion depends on the relation between initial 
velocity and ac force parameters (amplitude, frequency and phase), thus 
generalizing the results of Olsen and Samuelsen \cite{olsam}
for the non dissipative case. In the presence of damping, we also showed 
within the non-relativistic approximation that kink dc motion is never 
possible. This conclusion is different from that of 
Olsen and Samuelsen \cite{olsam}, who said ``similar [to the undamped case]
results are obtained'', not a very explicit statement; it is also
contrary to the claims of Bonilla and Malomed \cite{boni}, 
and finally it confirms the results of Cai {\em et 
al.} \cite{cai}. Numerical simulations 
fully reproduced the predictions of the approximate calculations establishing
their validity to a high degree of accuracy. In the present paper we 
considerably extend the study reported in \cite{anniu} in a systematic way, 
by analyzing i) the relativistic case for kinks, and ii) the effects of pure 
ac driving on breathers, that had not been studied so far. Our main conclusion
will be that the application of external ac driving forces 
can cause the dc motion of sG kinks (for the whole range of initial 
velocities) depending on their phase; for breathers, we show that in the 
absence of damping they can be self-sustained up to a certain threshold of
the force intensity, at which they break up into kink-antikink pairs. 
We also prove that combining the damping effect with the  
ac force the dc motion of sG kinks becomes asymptotically 
an oscillatory one.  In Sec.\ \ref{sec-cc}, we use a collective coordinate 
approach to obtain the equations of motion for the collective   
variables finding that the 
mean velocity of the kink (antikink) and the frequency of 
the breather are functions of the parameters of the ac force. 
We compare our analytical results by means of numerical simulations in 
Sec.\ \ref{sec-nr} and verify that they are in an excellent agreement. 
Finally, Sec.\ \ref{sec-c} summarizes our findings and conclusions, 
and presents a brief discussion about 
some other systems in which the 
phenomenon described in this paper can also appear. 
The paper closes with two appendices, one devoted to the proof of a 
mathematical intermediate result needed in Sec.\ \ref{sec-cc}, and the
other analyzing the effect of another dissipative term (with 
constant driving), $u_{xxt}$, of 
particular significance for Josephson junction dynamics, to show the 
power of our technique (see Sec.\ \ref{sec-cc}) 
to solve the relativistic problem.

\section{Collective coordinate approach}
\label{sec-cc}

\subsection{Kink dynamics}
\label{sub-k}

We begin our study by addressing the problem of kink dynamics, 
considering Eq.\ (\ref{ecua1}) when there is a single kink present in the
system with initial velocity $u(0)$, $t=0$ being taken as the time at which
the ac force is switched on. In order to study the subsequent kink evolution 
governed by Eq.\ (\ref{ecua1}), we apply the collective coordinate 
perturbation theory in the version introduced by McLaughlin and Scott 
\cite{McL}.  If $\beta$ and $\epsilon$ are small parameters we may assume 
that the solution of Eq.\ (\ref{ecua1}) has the same form as the unperturbed 
sG equation [Eq. (\ref{ecua1}) with $\beta=\epsilon=0$], except that now we 
allow $x_o(t)$,$X_{0}(t)$ and $u(t)$ to be functions of time, so that 
\begin{eqnarray}
\label{ecua2a}
\phi(x,t) & = & 4 \>{\rm arctan}\left( \exp\left[\pm \frac{x-x_{0}(t)-
X_{0}(t)}{\sqrt{1 -u^{2}(t)}}\right]\right), \\
X_{0}(t) & = &  \int_{0}^{t} {u(t')} dt',
\label{ecua2}
\end{eqnarray}
where the positive (negative) sign corresponds to a kink (anti-kink) solution. 
We note that $X(t) \equiv x_o(t)+X_{0}(t)$ [$u(t)$] 
has the meaning of the position (velocity) of the center of the
soliton, and that the main assumption
underlying this approximation is that radiation effects induced by the 
perturbation are neglected. This will be verified {\em a posteriori} by 
comparing with the numerical simulations.

With this {\em Ansatz}, we proceed much in the same way as McLaughlin and
Scott \cite{McL}: we first compute the variation of the energy and momentum 
of the unperturbed sine-Gordon system due to the perturbation, and applying
them to Eqs.\ (\ref{ecua2a}) and (\ref{ecua2}) obtain those variations in 
terms of our unknown functions. Imposing that it is consistent with the 
result of computing the variation of energy and momentum from the kink 
expression with constant parameters and making them functions after the 
calculation (see \cite{McL} for details; see also \cite{yo} for other 
versions of the collective coordinate technique), we arrive at the 
following equations for the position and velocity of the perturbed kinks 
(the $\pm$ sign corresponds to a kink or an anti-kink,
respectively):
\label{allecua3}
\begin{eqnarray}
\label{ecua3a}
\frac{du}{dt}&=&-\frac{1}{4} (1-u^{2}) [\pm \pi \sqrt{1-u^{2}} \epsilon f(t) 
+ 4 \beta u ], \\
u(t=0)&=&u(0); \nonumber\\
\label{ecua3b}
\frac{dX}{dt}&=&u(t),  \\
X(t=0)&=&X(0). \nonumber 
\end{eqnarray}
(equivalently, Eq.\ (\ref{ecua3b}) could written as $x_o(t) = const$).
For the dissipation-free case [$\beta=0$ in Eqs.\ (\ref{ecua1}) and 
(\ref{ecua3a})], Eq. (\ref{ecua3a}) can be exactly solved, yielding
\begin{eqnarray}
\label{ecua44}
u(t) & = & F[u(0),\delta_0]/\left(1+F[u(0),\delta_0]^2\right)^{1/2}, \\ 
F[u(0),\delta_0]& \equiv &  \frac{u(0)}{\sqrt{1-u^{2}(0)}}\pm\nonumber \\ 
& & \pm {\frac{\pi \epsilon}{4 \delta}}
[\cos(\delta t +\delta_{0})-\cos(\delta_{0})].
\label{ecua4}
\end{eqnarray}
Once we have obtained the velocity, we need to solve 
Eq. (\ref{ecua3b}) to find the kink trajectory. Unfortunately, it is not 
possible to find an analytical solution, so in order to get some insight 
about it we restrict ourselves to the non-relativistic limit, i.\ e., 
$|u(0)| \ll 1$, and imposing $|\pi\epsilon/4\delta| \ll 1$, 
(which is no loss of generality as we have already assumed $\epsilon \ll 1$)
we can find an approximate solution for the position of the center of the 
kink, $X(t)$, that reads
\begin{eqnarray}
X(t) & \simeq & X(0) + [u(0) \mp \frac{\pi \epsilon}{4 \delta} \, 
\cos(\delta_{0})]\>t\pm \nonumber \\
& &\pm \frac{\pi \epsilon [\sin(\delta t + \delta_{0})-
\sin(\delta_{0})]}{4 \delta^{2}}.
\label{ecua5}
\end{eqnarray}
 
Equation (\ref{ecua5}) tells us what the kink trajectory is, at least while
its velocity is not large. Keeping in mind this caveat, it is evident from 
the above expression that only for the ``resonant'' velocity
\begin{equation}
u(0) = \pm [\frac{\pi \epsilon}{4 \delta} \cos(\delta_{0})],
\label{ecua6}
\end{equation} 
the linear term will vanish and the motion of the kink (or antikink) 
will be oscillatory; for any other velocity values (fixed $\delta$, 
$\delta_{0}$ and $\epsilon$) the ac driving force induces a dc motion 
of sG kinks, even for the kinks initially at rest. This result contains
those found by Olsen and Samuelsen \cite{olsam} (also obtained in the 
non-relativistic approximation, by the way)  as particular cases for
two choices of $\delta_0$. 

So far, we have developed a collective coordinate theory in the 
non-relativistic limit. This have already provided information on 
the kink dynamics, but, in addition, we can benefit from that to 
find the unrestricted relativistic result, by 
virtue of the following fact: It can be proven (see Appendix I) 
from equations (\ref{ecua3a}) and (\ref{ecua4}) that the relation  
\begin{equation}
\int_{0}^{T}{F[u(0),\delta_0]} dt=0,   
\label{ecua7}
\end{equation} 
is a necessary and sufficient condition for the motion to be oscillatory 
with period $\displaystyle T=2 \pi/\delta$. 
Hence, integrating (\ref{ecua7}) we find that kinks 
will oscillate if and only if  
\begin{equation}
\frac{u(0)}{\sqrt{1-u^{2}(0)}} = 
\pm [\frac{\pi \epsilon}{4 \delta} \cos(\delta_{0})]. 
\label{ecua8}
\end{equation}

Let us now discuss this condition. To begin with, 
the condition (\ref{ecua8}) involves four parameters, 
so we can fix, say, $\delta$, 
$\epsilon$ and $u(0)$ and then use (\ref{ecua8}) to find the 
value of $\delta_{0}$ such that the motion becomes oscillatory.  
Second, the only change with respect to the approximate  
condition (\ref{ecua6}) is the appearance of the Lorentz factor 
$\gamma={\sqrt{1-u(0)^{2}}}$ in Eq. (\ref{ecua8}), which vanishes in the small 
velocity limit, i.\ e., the result (\ref{ecua6}) is recovered. 
Equation (\ref{ecua8}) is very important, because it implies that 
(within the collective coordinate approach), any ac force applied on 
a kink will induce its dc motion with a velocity directly related to 
their relative phase [vanishing only for a very special choice of
that parameter, $\delta_0$]. Our perturbative calculation coincides
with that of Olsen and Samuelsen in the limit and special cases studied
by them, thus generalizing their results in a comprehensive manner. 
We postpone the numerical validation of our findings to Sec.\ \ref{sec-nr}.

After dealing with the undamped case and arriving to an expression valid
for all kink velocities, let us now turn to the damped problem, including
a damping term ($\beta \ne 0$) in the ac driven sG equation.
We will now demonstrate that dc motion of sG kinks is not possible in
this case. Indeed, in our previous work \cite{anniu} we analized the 
effects of damping ($\beta \ne 0$) and ac force non-relativistic kinks 
and found that kinks exhibit dc motion except for a transient, after 
which they reach a final oscillatory state around a point whose location
depends on the initial conditions. However, nothing was said there about
the general case, with no limits on the initial velocity, so we need to
see what is the behavior of relativistic kinks by looking back again at
Eq.\ (\ref{ecua3a}). The result found above for the undamped case 
suggests to use $y(t)=u(t)/\sqrt{1-u(t)^{2}}$ as a new variable; by doing
so, Eq.\ (\ref{ecua3a}) can be linearized and the analytical solution 
for the velocity computed. We thus arrive at 

\begin{eqnarray}
\label{ecua99}
u(t) & = & \frac{r(t)}{\sqrt{1+r(t)^2}},\\
r(t) & \equiv & \bar{c} \, \exp(-\beta t) \mp 
\frac{\pi \epsilon}{4 (\beta^{2} + \delta^{2
})} \times \nonumber \\
\label{ecua999}
& & \times [\beta \sin(\delta t + \delta_{0}) - \delta \cos(\delta t +
\delta_{0})],\\
\bar{c}&=&\frac{u(0)}{\sqrt{1-u(0)^{2}}} \pm \frac{\epsilon \pi}
{4 (\beta^{2} + \delta^{2})}\times\nonumber\\ 
& &\times [\beta \sin(\delta_{0})- \delta \cos(\delta_{0})].
\label{ecua9}
\end{eqnarray} 
{}From Eqs.\ (\ref{ecua99},\ref{ecua999},\ref{ecua9})
it can be easily seen that the asymptotic 
behavior of $u(t)$ is given by
\begin{eqnarray}
\bar{u}(t) & = & \frac{\bar{r}(t)}{\sqrt{1+\bar{r}(t)^2}},\\
\bar{r}(t) & = & 
\pm \frac{\pi \epsilon}{4 (\beta^{2} + \delta^{2
})} \times \nonumber \\ & \times &
[\delta \cos(\delta t + \delta_{0}) - \beta \sin(\delta t + \delta_{0})].
\label{ecua10}
\end{eqnarray}
Even if we cannot explicitly solve 
Eq.\ (\ref{ecua3b}) when $u(t)$ is given by 
Eqs.\ (\ref{ecua99},\ref{ecua999},\ref{ecua9}) for the 
trajectory $X(t)$, the above findings are all we need to prove (always 
within the collective coordinate formalism) our previous claim that 
kink dc motion is asymptotically forbidden. To this end, it is enough 
to show that for $t_{0}$ and $t$ large enough, 
$X(t) \sim \bar{X}(t)=\bar{X}(t_{0})+
\int_{t_{0}}^{t} {\bar{u}(t')} dt'$, is a periodic function. 
This is equivalent to proving that 
\begin{equation}
\int_{t}^{t+T}{\bar{u}(t^{\prime})} dt^{\prime}=0;
\label{ecua11}
\end{equation}
A straightforward calculation shows that the condition 
(\ref{ecua11}) is always true, giving 
grounds to our conclusion that kinks can not 
exhibit dc motion in the damped, ac driven sG equation. This is yet another
proof that the arguments of Bonilla and Malomed \cite{boni} supporting 
the opposite conjecture were not correct. 

\subsection{Breather dynamics}
\label{sub-b}

Following our study for kinks, we have carried out an analysis of the 
effect of ac driving over the breather solutions, based again 
on the collective coordinate approach (see details in 
\cite{McL,bismc}). We concern ourselves mostly with the undamped case, as 
the damped problem has been already carefully discussed in 
\cite{lomsam1,lomsam2}. The collective coordinate {\em Ansatz} we 
will use for breathers is 
\begin{equation}
\phi(x,t) = 4 \arctan\left( \frac{g(t)}
{\cosh[x \hspace{0.1cm}\omega(t)]}\right),     
\label{ecua12}
\end{equation}
where the breather frequency is $\Omega(t)=\sqrt{1-\omega^{2}(t)}$. 
This {\em Ansatz} is assumed to be valid when   
$\beta$ and $\epsilon$ are small.
In this case the collective variables are the two functions 
$\omega(t)$ and $g(t)$, which can be found solving a system 
of differential equations equivalent to that in Eqs.\ 
(\ref{ecua3a}, \ref{ecua3b}) for the kinks. The procedure to find
the equations is the same and yields 
\begin{eqnarray}
\label{ecua13a}
\frac{d\omega}{dt} & = & \frac{\pi \epsilon}{4} \frac{f(t)}
{\sqrt{1 + g^{2}(t)}} \times \nonumber\\
& &  \times \cos\left\{\arcsin[\frac{\sqrt{1-\omega^{2}(t)}}{\omega(t)} 
g(t)]\right\}, \\
\omega(t=0)& = &\omega_{0} \nonumber\\
\frac{dg}{dt} & = & \left\{\omega(t) + \frac{\pi \epsilon}{4 \omega(t)} 
f(t) \mbox{arcsinh}[g(t)]\right\}  \times \nonumber \\ 
\label{ecua13}
& & \times \cos(\arcsin[\frac{\sqrt{1-\omega^{2}(t)}}{\omega(t)} g(t)]),  \\
g(t=0)& = &0.  \nonumber
\end{eqnarray}
We have not been able to solve the above equations. Nevertheless, they 
are still useful in order to predict the breather stability conditions,
as we now show. 
If $\omega(t) \ne 0$, the system of Eqs.\ 
(\ref{ecua13a},\ref{ecua13}) leads us to  
\begin{eqnarray}
\frac{1}{\omega} \frac{d\omega}{dt} & = &
\frac{\pi \epsilon}{4 \omega^{2}} \frac{f(t)} {\sqrt{1 + g^{2}(t)}}
\times \nonumber \\ 
& & \times \frac{1}{(1 + \frac{\pi \epsilon}{4 \omega^{2}} 
f(t) \mbox{arcsinh}[g(t)])}\frac{dg}{dt}. 
\label{ecua14}
\end{eqnarray}
Straightforward algebraic manipulations of this equation yield
\begin{equation}
\frac{d}{dt}\ln[\omega(t)] = 
\frac{d}{dt} \ln[1 + \frac{\pi \epsilon}{4 \omega^{2}(t)} 
f(t) \mbox{arcsinh}(g[t])]. 
\label{ecua15}
\end{equation} 
where terms of order $\epsilon^{2}$ 
and $\epsilon \delta$ have been neglected. 
The advantage of this expression is that we can now integrate it; 
integration and subsequent expansion of $\omega(t)$ 
around $\omega_{0}$ in powers of $\epsilon$ allow us to write
(up to orders $\epsilon^2$ again)
\begin{eqnarray}
\label{ecua16}
\omega(t) & = &  \omega_{0} [1 + \sigma(t)], \\ 
\label{sigma}
\sigma(t) & \equiv & \frac{\pi 
\epsilon}{4 \omega^{2}_{0}} 
f(t) 
\mbox{arcsinh}\left[\frac{\omega_{0}\sin(\sqrt{1 - 
\omega^{2}_{0}} t)}{\sqrt{1 - \omega^{2}_{0}}} 
\right]. 
\end{eqnarray}

In order to proceed, we have to recall the following property:
breather solutions can only exist provided that 
$0 < \omega(t) < 1$ (which implies that the breather energy is
bounded above and below, $ 0 < H_{bre}(t) < 16$, kinks having
energy equal to or larger than 8 in our dimensionless units). Having this 
in mind, we can find a condition for the transformation of a breather 
into a $k\bar{k}$ pair, that turns out to be 
 
\begin{eqnarray}
\epsilon_{thr} & = & \min\left(\epsilon_{1},\epsilon_{2}\right),
\hspace*{0.5cm} \mbox{ if } M>0 \mbox{ and } m<0, \nonumber\\
\label{ecua17}
\displaystyle \epsilon_{thr} & = & \epsilon_{2}, \hspace*{0.5cm} 
\mbox{ if } M \le 0, \\
\epsilon_{thr} & = & \epsilon_{1} \hspace*{0.5cm}
\mbox{ if } m \ge 0, \nonumber
\end{eqnarray}
with 
\begin{eqnarray} 
\label{ecua17a}
\epsilon_{1} &=& \frac{4 \omega_{0} (1 - \omega_{0})}{\pi M},\\
\label{ecua17b}
\epsilon_{2} &=& - \frac{4 \omega_{0}^{2}}{\pi m},
\end{eqnarray}
$m$ and $M$ being the minimum and maximum respectively of the function  
$\sigma(t)$ [see Eq.\ (\ref{sigma}) above]. 

The physical meaning of the somewhat cumbersome result is the following: 
when sG breathers are perturbed by an ac force there always exists a
value of $\epsilon$, $\epsilon_{th}$, such that perturbations smaller
than the threshold allow breathers to exist
with modulated frequency $\Omega(t)=\sqrt{1-\omega^{2}(t)}$, where   
$\omega(t)$ is the function (\ref{ecua16}); on the other hand, when 
$\epsilon>\epsilon_{thr}$ the breather transforms into a kink-antikink 
pair or decomposes into a radiation, depending of the parameters of ac force 
and its frequency. Therefore, we arrive at our major conclusion regarding 
(undamped) breather dynamics, very different from the kink analysis in 
the previous subsection: Breathers are only stable for a limited range 
of force intensities, and in that range the effect of the driving is 
to modulate their frequency as 
discussed above. As before, we will have to compare this prediction to 
numerical simulations to verify its validity. 

\begin{figure}
%\vspace{1cm}
\epsfig{file=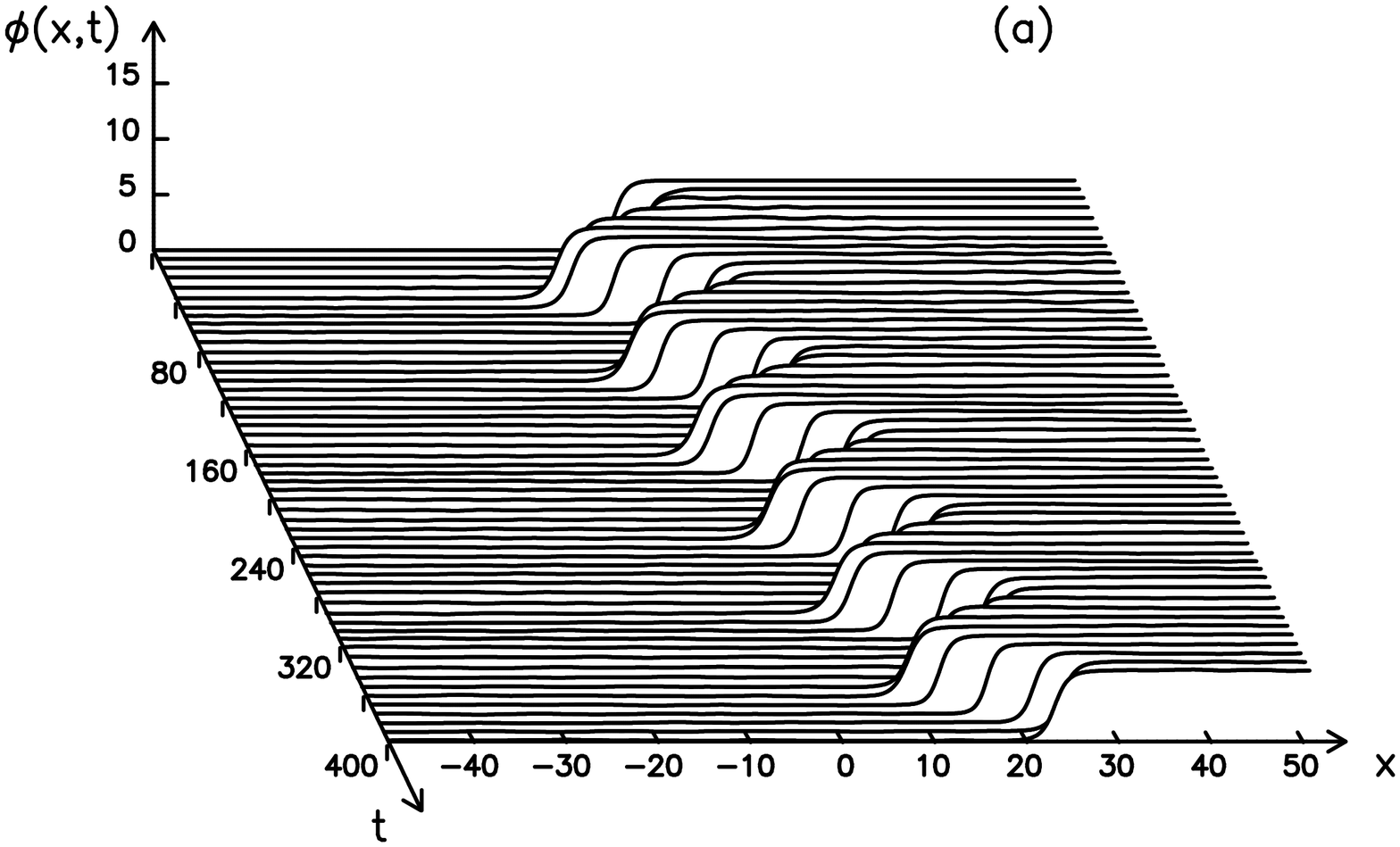, width=2.5in, angle=-90}

\vspace{-2cm} \epsfig{file=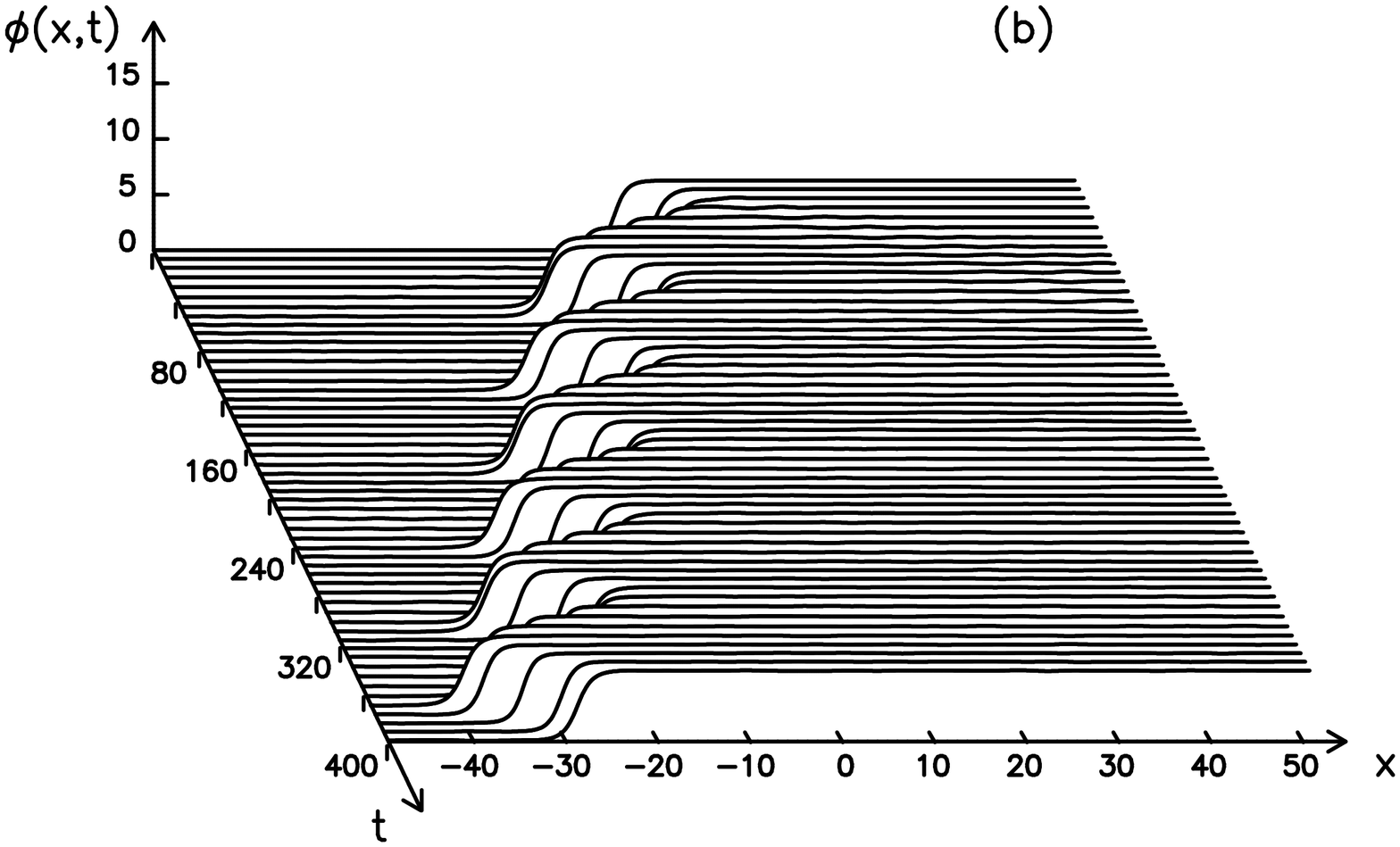, width=2.5in, angle=-90}

\vspace{-2cm} \epsfig{file=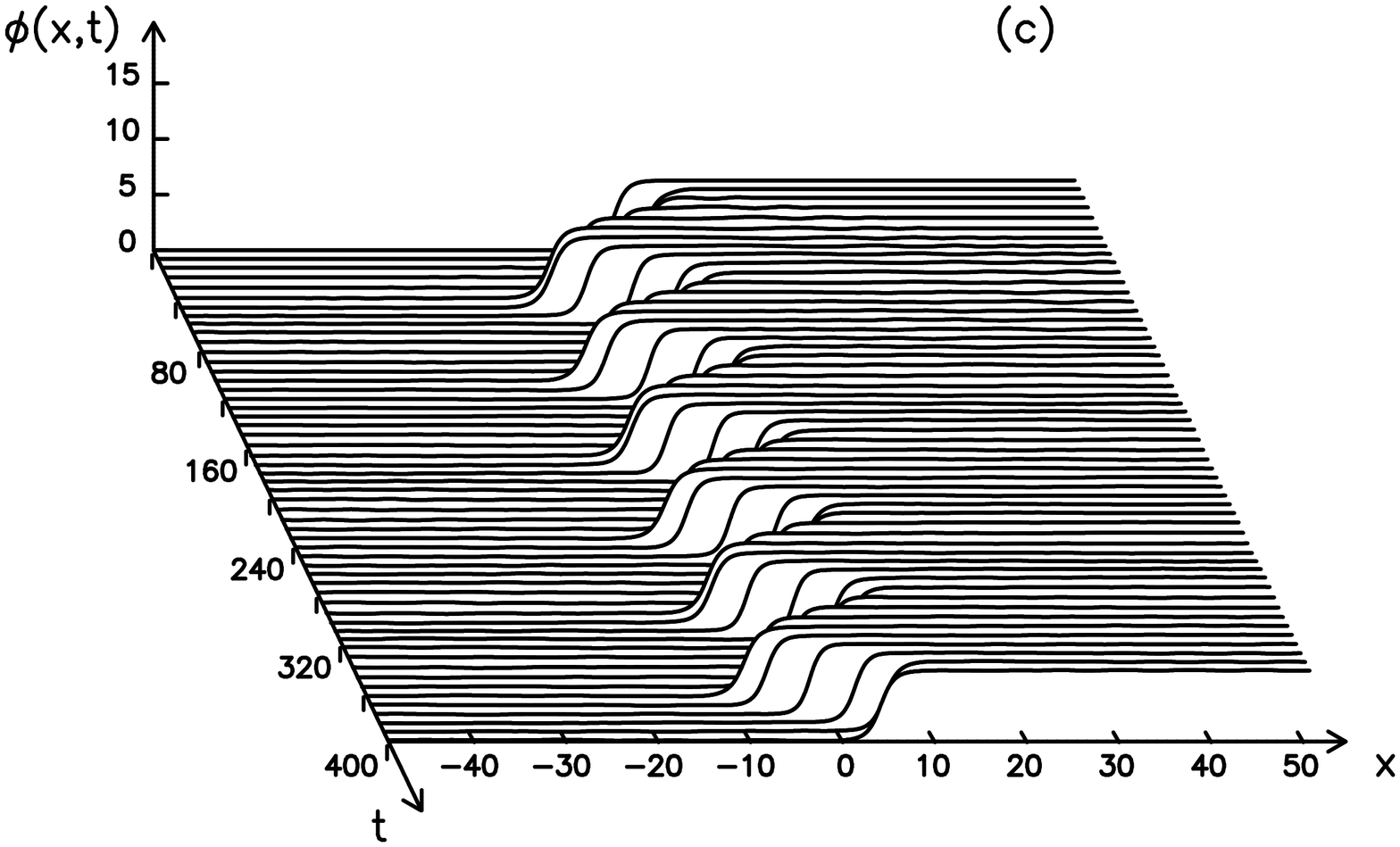, width=2.5in, angle=-90}

\vspace{-2cm} \caption{Verification of collective coordinate predictions in the
absence of dissipation. Simulation 
starts from a kink with initial velocity $u(0)=0.5$ located at 
$X(0)=0$, and subject to an ac force given by $0.1
\sin(0.1 t + \delta_{0})$. (a) $\delta_{0}=0.75$,  
(b) $\delta_{0}=0$; notice that 
the direction of motion is opposite in both cases.
(c) $\delta_{0}=0.6$, critical value exhibiting oscillatory motion.} 
\label{graf1c}
\end{figure}
\begin{figure}
\epsfig{file=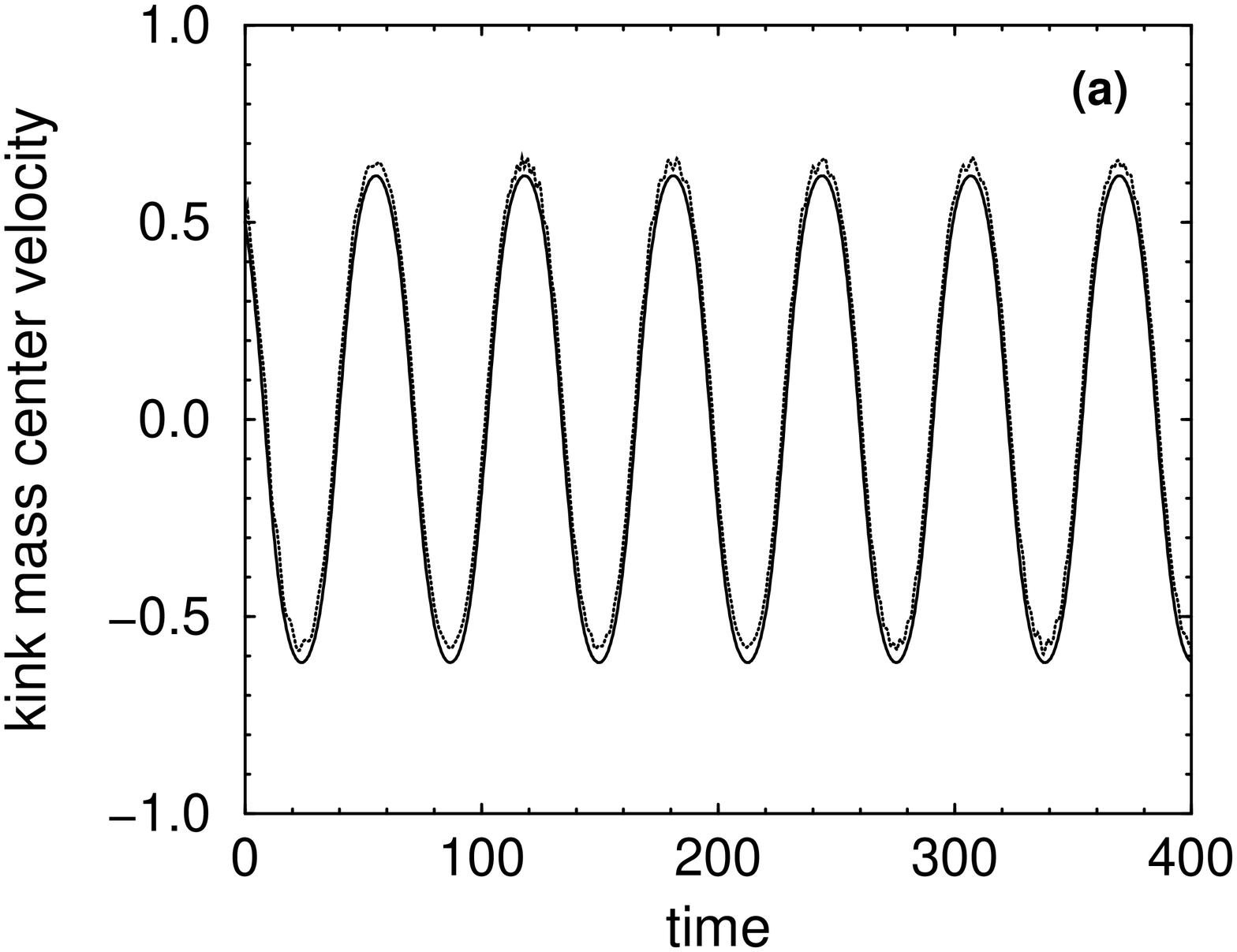, width=2.5in, angle=-90}
\epsfig{file=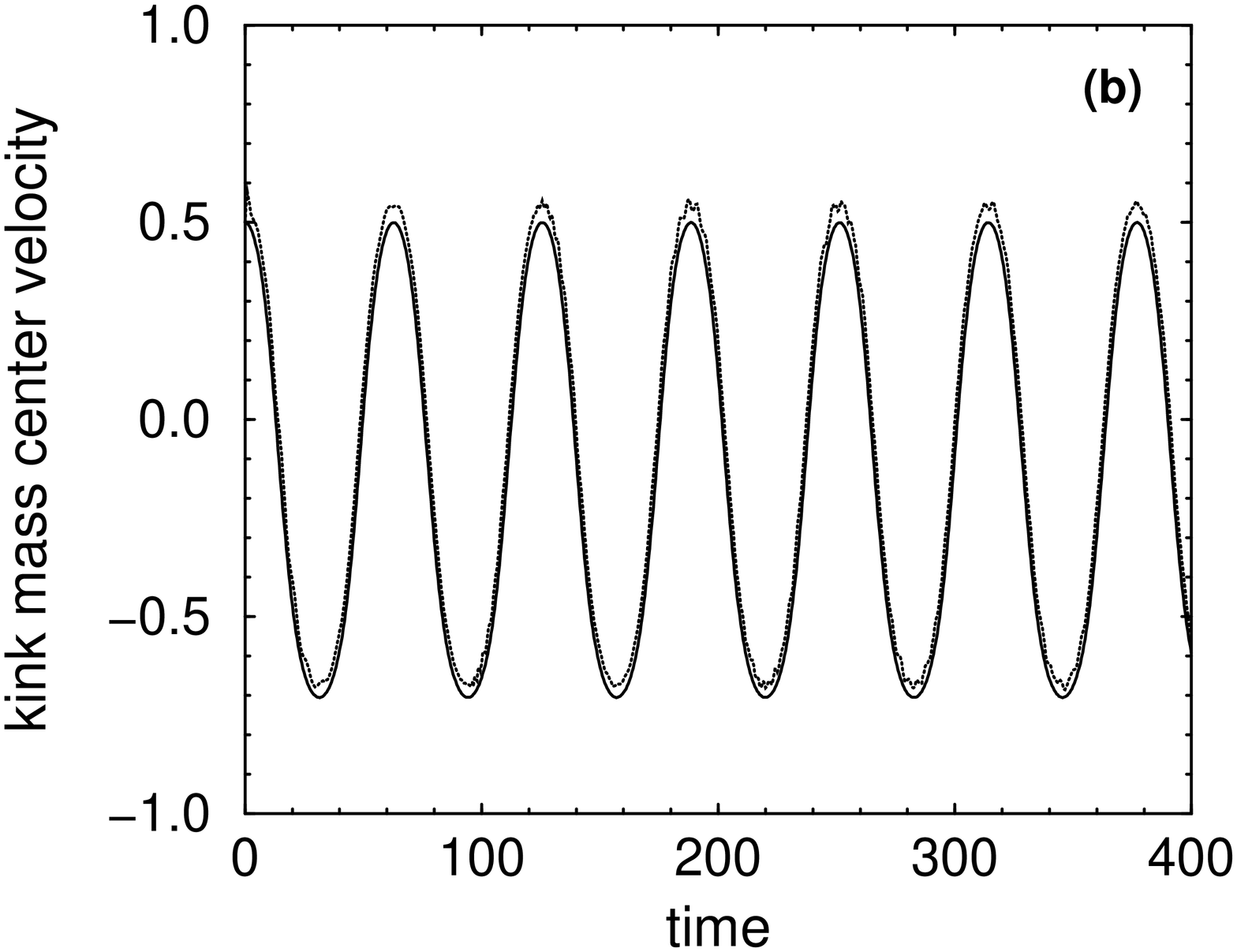, width=2.5in, angle=-90}
\epsfig{file=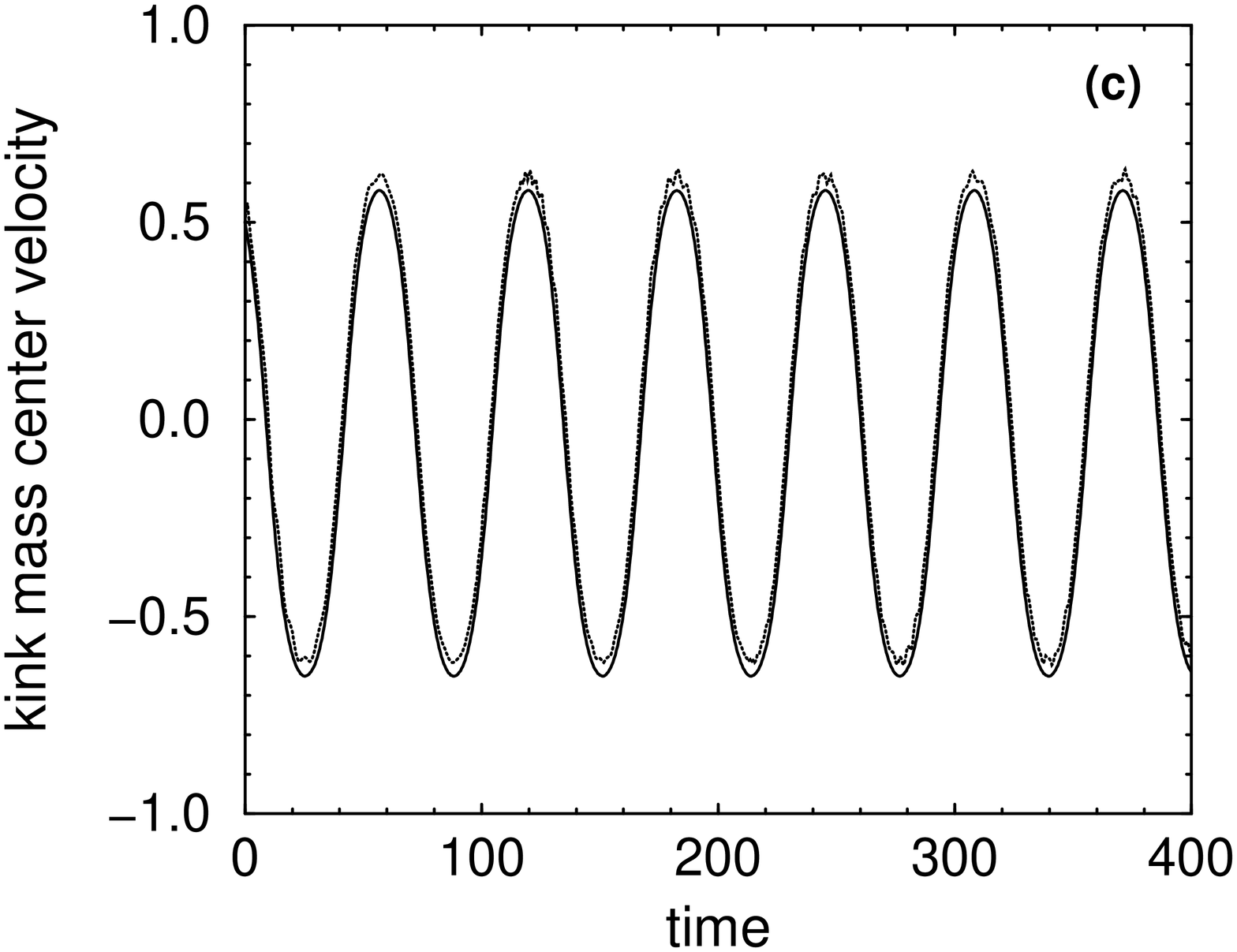, width=2.5in, angle=-90}
\caption{Time evolution of the velocity of kink center mass. 
Parameter values as in Fig. 1. (a), (b) and (c) respectively. 
The solid curve shows the value of velocity, predicted by the 
collective coordinate perturbation [Eqs. (\ref{ecua44},\ref{ecua4})], and the 
dotted line is the numerical integration of [Eq. (\ref{ecua1})]}. 
\label{velocity}
\end{figure}

\section{Numerical results}
\label{sec-nr}

Although the collective coordinate approach has been already tested very 
many times \cite{yo} and is by now a standard technique to deal with 
soliton-bearing equations, the analytical expressions and the conclusions
we have drawn from them should not be taken as valid before comparison with
detailed and accurate numerical simulation of the full partial differential
equation (\ref{ecua1}). To this end, we have numerically integrated it by 
means of a standard fourth order Runge-Kutta algorithm \cite{nrecipies} 
with initial conditions given by an unperturbed
sine-Gordon kink or breather with unrestricted initial 
velocity $u(0)$, and boundary conditions
$\phi_x(L=\pm 50,t) = 0$. We have carefully checked that our results 
did not depend on the choice of spatial and temporal steps, or the 
simulated system size (which can affect the simulations through the 
generation of spureous radiation). 

We first address the validation of the results for kinks, beginning with
the undamped case. In Fig.\ \ref{graf1c} we show the time evolution 
of a sG kink with initial velocity $u(0)=0.5$ (which is certainly far 
from being small)
for different phases of the ac force $\delta_{0}$ and $\epsilon=0.1$, 
$\delta=0.1$. From Eq.\ (\ref{ecua8}) we know that the kink (with initial 
velocity $u(0)=0.5$) will oscillate only if $\delta_{0}=0.745$. Numerically,
we searched for the critical phase of driving force and 
found $\delta_{0}=0.6$ (see Fig.\ \ref{graf1c}c), i.e., the accuracy 
between these results it is of order of $0.1$. This is a very good prediction
if we consider the value of the velocity and the value of $\epsilon$ (indeed,
the accuracy is of the order of $\epsilon$ as should be expected). Note that 
there is practically no radiation visible in the simulations, which is in 
agreement with the correctness of the use of collective coordinate 
techniques. In connection with this, 
another interesting remark is that background motion, i.e., 
motion of the wings of the soliton induced by the ac force, is 
not visible in the simulations; this makes our approach very suitable for
this problem, and as accurate as that of Olsen and Samuelsen \cite{olsam}
who introduced a term accounting for this negligible effect. 
We note that smaller initial velocities compare even better with our 
theoretical results. To appreciate to a larger extent the accuracy of our 
approach, Fig.\ \ref{velocity} compares the analytical prediction 
for the velocity values [Eqs.\ (\ref{ecua44},\ref{ecua4})] with 
the velocity of the kink center of mass, obtained by 
integrating numerically Eq.\ (\ref{ecua1}). In these figures the parameters 
of the perturbed sG equation were chosen as in Fig.\ \ref{graf1c}. We see from 
Fig.\ \ref{velocity} 
that the agreement is excellent, in spite of the fact that the
prediction of the critical velocity is not correct by a factor $\epsilon$. 
This is due, on one hand, to the fact that slight deviations from simmetry around 
$u(t)=0$ have dramatic consequences in the kink motion (note, e.\ g., Fig.\ 
\ref{velocity}c, the plot
for the critical velocity, where the small discrepancy between the theory and the
simulation is always in the same direction, towards negative $u$, corresponding to
the theoretically predicted dc motion to the left), and, on the other hand, to the 
difficulties involved in numerically determining the critical velocity by checking 
that the kink just oscillates; this, among other things, depends on the integration
time and on the time step. Hence, we believe that the correctness of our computations
is much better appreciated by looking at the function $u(t)$ than simply verifying 
the critical velocity value.

\begin{figure}
%\vspace{1cm}
\epsfig{file=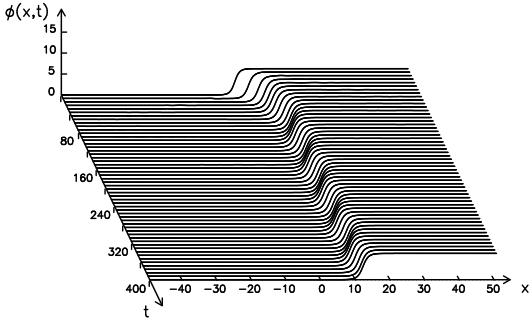, width=2.5in, angle=-90}

\vspace{-2cm} \caption{Verification of collective coordinate predictions with
dissipation. This is one example of soliton motion, initially with 
velocity $u(0)=0.5$, and  
subject to an ac force given by $0.01 \sin(0.1 t + \delta_{0})$, 
with $\delta_{0}=0.1$ and $\beta=0.05$.} 
\label{graf2}
\end{figure}

\begin{figure}
\epsfig{file=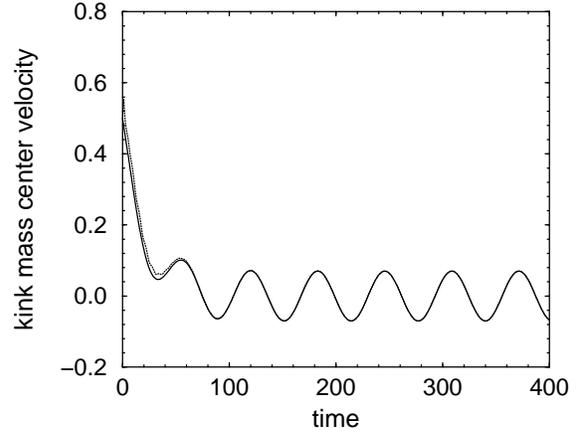, width=2.5in, angle=-90}
\caption{Time evolution of the velocity of kink center mass. 
Parameter values as in Fig. 3.  
The solid curve is the analytical prediction
[Eqs.\ (\ref{ecua99},\ref{ecua999},\ref{ecua9})]
and the dotted curve is the numerical integration of [Eq. (\ref{ecua1})].} 
\label{velocitydis}
\end{figure}

We cannot close our study about  sG kinks without providing numerical evidence that,
as predicted by our analytical calculations above,
it is not possible to induce dc motion on sG kinks by means of an 
ac force in the presence of damping.
Figure \ref{graf2} shows an example: a kink, initially moving with 
velocity $u(0)=0.5$, perturbed by ac force $0.01 \sin(0.1 t + 0.1)$ and a 
damping term $\beta=0.05$, is stopped and ends up describing small amplitude
oscillations around a finite region in space. This is but one example, although
we must say that we have verified that our prediction is correct in very many 
other cases, with different choices of parameters, finding always an excellent
agreement between analytics and numerics. As in the undamped case, Fig.\ 
\ref{velocitydis}, depicts the predicted and simulated motions of the kink 
center, exhibiting a perfect agreement between both. It is of course reasonable
that the collective coordinate approach works better for the damped case, because
if any radiation arises in the system, the damping wipes it away, leaving the 
kink as the only excitation present in the system, as the approximation requires. 

\begin{figure}
\epsfig{file=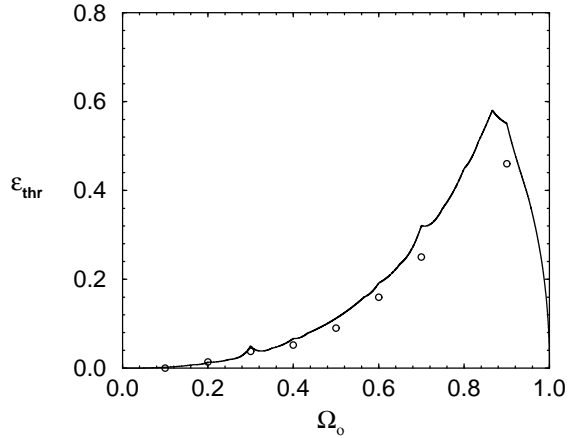, width=2.5in, angle=-90}
\caption{Illustration of the threshold value of $\epsilon$ vs the initial 
frequency $\Omega_{0}$. 
Solid line represent the values given by the perturbation 
analysis and the points are the values, obtained by numerical simulation 
of perturbed sine-Gordon equation. The frequency and the phase of ac force 
are $\delta=0.1$, $\delta_{0}=0$ respectively.} 
\label{graf3}
\end{figure}

\begin{figure}
\epsfig{file=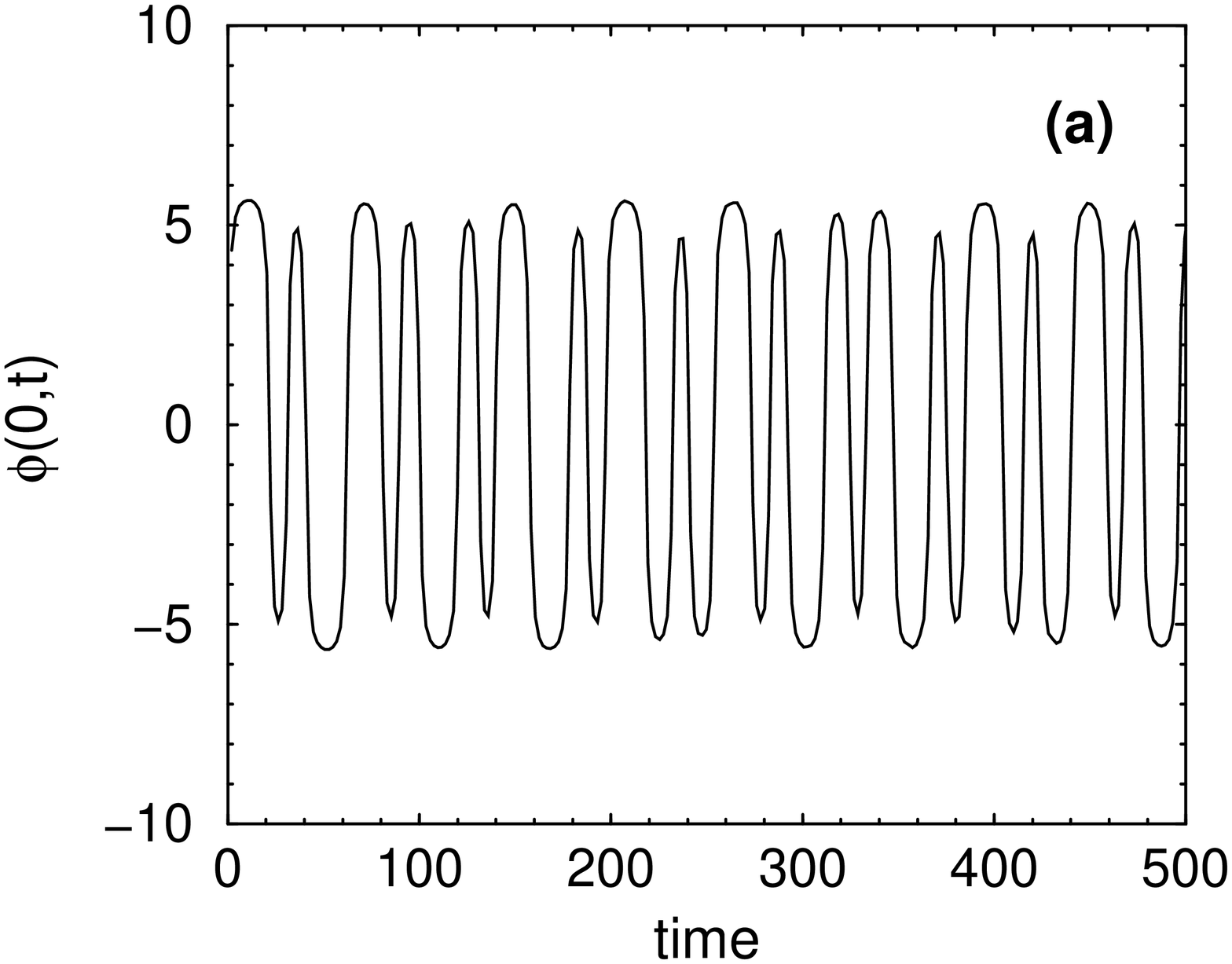, width=2.5in, angle=-90}
\newpage
\epsfig{file=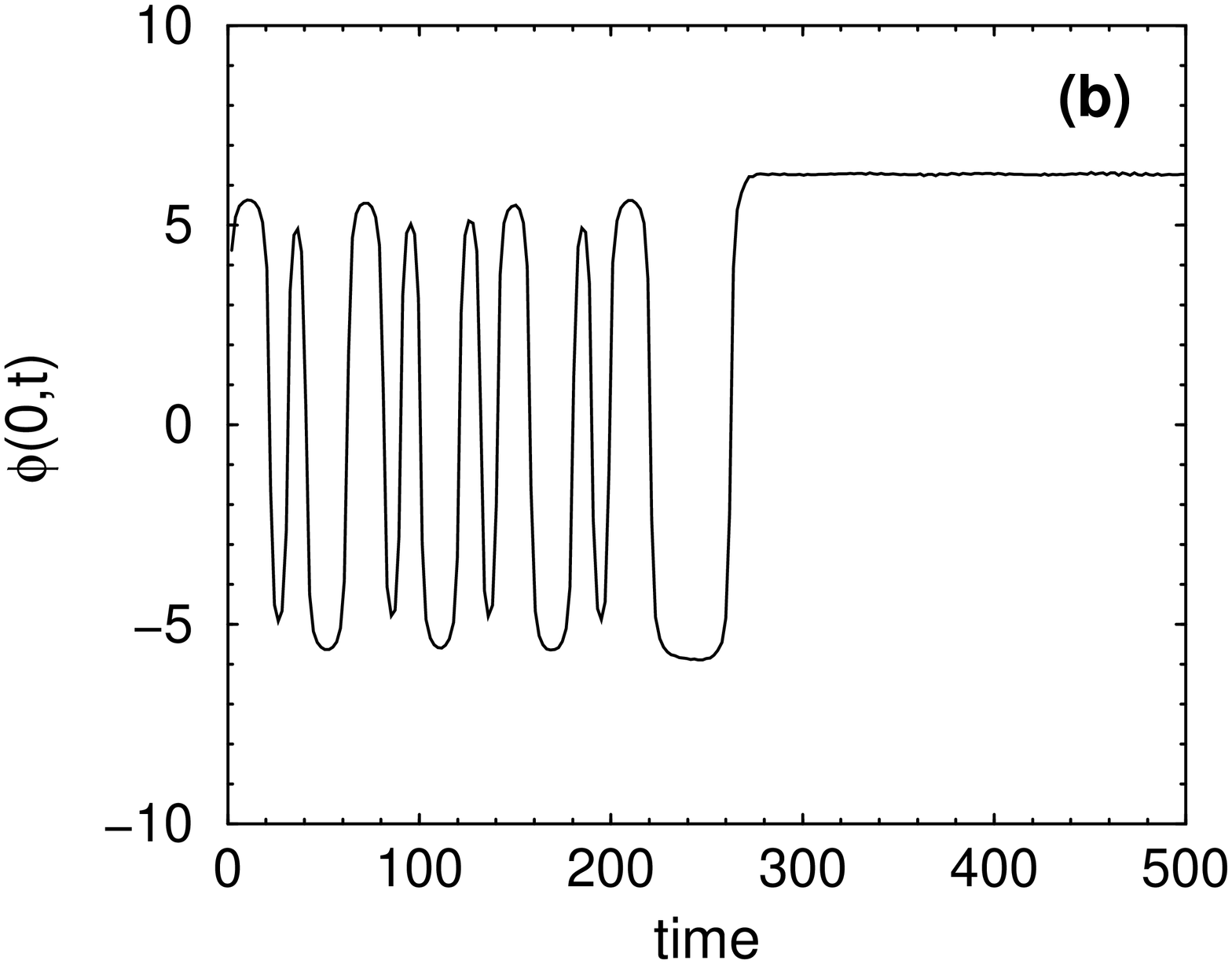, width=2.5in, angle=-90}
\caption{(a) The breather with modulated frequency for the value of 
$\epsilon=0.0148$.(b) If $\epsilon=0.0149$ the breather break into $k\bar{k}$ 
pair. The others parameter values are $\Omega_{0}=0.2$, $\delta=0.1$, 
$\delta_{0}=0$. Compare with the analytical threshold  $\epsilon$ 
($\epsilon_{th}=0.0135$), obtained in Eqs.\ 
(\ref{ecua17},\ref{ecua17a},\ref{ecua17b}).} 
\label{graf4}
\end{figure}

Finally, for the breather case we present in 
Fig.\ \ref{graf3} a comparison between the analytical and numerical
threshold values of $\epsilon$ 
as a function of the initial frequency of the breather $\Omega_0$.
We consider only the undamped case, $\beta=0$. 
We have observed in our simulations the phenomenon predicted by our analytical
calculations: If the amplitude of the ac force exceeds the threshold value of 
$\epsilon$, the initial breather decomposes into a $k\bar{k}$ pair. As depicted
in Fig.\ \ref{graf3}, the collective coordinate result predicts very well the 
threshold amplitude in all the range of breather frequencies, even for rather 
large values of such amplitude, where the perturbative approach might not work
in principle.
We note that,
for driving frequencies close to 1 (the lower edge of the phonon band of the 
sG system), the breather can as well decay into radiation, but we have not pursued
this process in detail in view of the difficulties involved in distinguishing 
breathers from linear radiation in this limit. Therefore, we concern ourselves 
with the decay into $k\bar{k}$ pairs, this being the reason we have not gone 
over $\Omega_0=0.9$ \cite{not} in our numerical tests. Figure \ref{graf4} shows an 
instance of our numerical determination of the threshold, for a 
breather with $\Omega_{0}=0.2$. We
can see that if $\epsilon=0.0148$ (Fig.\ \ref{graf4}a) the frequency of 
the breather is modulated, but it remains a breather, as the oscillations 
of the field value at $x=0$ clearly show (of course, we have looked at the
simulation over the whole spatial interval to confirm this). On the contrary,
if we increase $\epsilon$ (for example, $\epsilon=0.0149$) then the breather 
breaks up
into a $k\bar{k}$ pair (Fig.\ \ref{graf4}b), a phenomenon which reveals itself
in the suppression of the oscillations at $x=0$ indicating that the breather
stops its periodic motion. We have thus verified the validity of 
the collective coordinate results for the breather case as well.

\section{Conclusions}
\label{sec-c}

In this paper, we have shown both by analytical (collective coordinate
theory) and numerical means the following three facts: i) Kink dc motion
is possible in the sG system when forced by a pure ac force, except for 
a very special choice of the initial velocity related to the driving phase
by equation (\ref{ecua8}); in this case, the kink center 
oscillates around its initial position; ii) kink dc motion induced by 
pure ac driving is never possible in the presence of damping, and iii) 
the undamped sG system forced by an ac driving allows the existence of 
breathers, whose frequency becomes modulated by the external one, up to
a certain critical value of the amplitude that depends on the driving 
frequency and on the breather parameters.  Above that amplitude, the 
breather decays to a kink-antikink pair. 

As already stated, this findings arise from a perturbative calculation of
the collective coordinate type, whose accuracy for very many soliton 
problems is well established \cite{yo}. In this respect, we want to emphasize
our main achievement in this study: We have proposed a change of variables 
which allows to linearize and fully solve the differential equation for 
the kink velocity without any restriction. We note that in previous works
\cite{olsam,anniu} only the non-relativistic limit could be dealt with in an 
approximate manner. It is clear that 
this analytical procedure can be useful in similar 
equations, an example of this being our treatment of the long Josephson 
junction equation, with a more realistic dissipative term and a constant
driving, in Appendix II. 
Therefore, having in mind the accuracy of the collective
coordinate approach, the fact that there were no further approximations 
involved in our calculations, and the numerical confirmation of our predictions,
we believe that our study settles down once and for all the question of 
kink motion in ac driven soliton bearing systems, particularly so for the
sG equation. 

{}From these results, which clarify the phenomenology of the ac driven 
sG equation, we can try to draw some more general conclusions on 
soliton-bearing problems perturbed by an ac force. The findings 
on discrete sG systems \cite{zaragoza} suggest that for our conclusions
to apply in other, different systems, it is necessary that the soliton 
or solitary wave under consideration does not have one or several internal
oscillation modes which can be excited by the driving force. To check for
the possible new phenomena or modifications of our conclusions coming from
the existence of those modes, we have carried out some 
preliminary research on the ac driven 
$\phi^{4}$ model. Generally speaking, the same conclusions hold, 
i.\ e., in the absence of dissipation ($\beta=0$) the kink 
(antikink) will exhibit oscillatory motion if 
\begin{equation} 
\frac{u(0)}{\sqrt{1-u^{2}(0)}} = 
\pm [\frac{3 \epsilon}{\sqrt{2} \delta} \cos\, 
(\delta_{0})], \nonumber
\end{equation}
whereas when dissipation is present we again find that kink dc motion is 
not possible. However, we have found numerical evidence that for values of 
the driving frequency close to the internal mode one, the kink behavior is
much more complicated, even chaotic. We are currently working on 
understanding this effect analytically, and tentatively we relate 
it to the resonance of the driving and the internal mode \cite{physd}. 
Similar resonance phenomena were observed for the sG kink in a harmonic
potential well, when the driving had a frequency close to the natural 
one of the kink in the well \cite{fernandez}, which reinforces our 
interpretation and expectations. 
This leads us to the conjecture that the results we have obtained for 
the sG problem can be true in general, provided as already said that there
are no resonances with other modes intrinsic to the excitation. Of course,
there also remains the question of drivings with frequencies above the 
phonon lower edge. We believe that our conclusions will still hold in 
that situation, in so far the presence of the external force does not 
cause an interaction between solitonic and radiation modes; it is to 
be expected that our calculations will be very accurate in the presence 
of dissipation, which will damp away any excited radiation. Finally, as
shown in Appendix II below, we want to stress that the predictions and 
mathematical results we have obtained in this work have direct application
in many physical systems, and among them in Josephson devices [more 
realistically described by Eq.\ (\ref{aecua1}) below]. It would be most
interesting to verify the possibility of dc motion of solitons induced 
by pure ac driving in real systems like that, more so in view of the 
potential applications such a rectification effect might have. Another
implication of our results has to do with soliton generation in systems
described by sG equations: Indeed, by driving the system with an ac 
force strong enough, we could break up thermally created breathers into 
kink-antikink pairs, which under the influence of the driving would 
separate moving towards opposite directions. This would be a very simple
and clean way of creating solitons in, e.g., Josephson devices. 
It thus becomes clear that
experimental work pursuing this and related questions, which 
is certainly amenable with the present capabilities, would be most 
useful in assessing the importance of our work.

\begin{acknowledgement}
We thank Renato \'Alvarez-Nodarse, Jose Cuesta,  Esteban Moro and 
Franz Mertens  for conversations on this work.
Work at GISC (Legan\'es) has been supported by CICyT (Spain) grant MAT95-0325
and DGES (Spain) grant PB96-0119.
\end{acknowledgement}

\section*{Appendix I}
\label{sec-aI}

In this appendix we prove the necessary and sufficient condition for the
sG soliton motion to be purely oscillatory in the presence of undamped 
ac driving. In the body of the paper we stated that this is so if and 
only if 
the condition (\ref{ecua7}), or equivalently, (\ref{ecua8}), takes place.

On the one hand, 
Substituting (\ref{ecua8}) in (\ref{ecua4}) and afterwards in 
(\ref{ecua44}) we obtain for the velocity 
function $u(t)$ the following expression:
\begin{equation}
u(t) = \frac{\pm {\frac{\pi \epsilon}{4 \delta}}
\cos(\delta t +\delta_{0})}{\sqrt{1+
\left[{\frac{\pi \epsilon}{4 \delta}}
\cos(\delta t +\delta_{0})\right]^{2}}},
\label{ecua19}
\end{equation}
which can be integrated, yielding the position of the kink center of 
mass $X(t)$:
\begin{equation}
X(t) = X(0) \pm \frac{1}{\delta} 
\arcsin\left(\frac{\pi \epsilon \sin(\delta t + \delta_{0})}
{\sqrt{16 \delta^{2} + \pi^{2} \epsilon^{2}}}\right).  
\label{ecua19a}
\end{equation}
Hence, it is evident from the above equation that
$X(t)$ is a periodic function with period $T={2 \pi/\delta}$. 

On the contrary, 
suppose that the kink center of mass 
\begin{equation}
X(t) = \int_{0}^{t} {u(t')} dt'
\end{equation} 
oscillates with period 
$T={2 \pi/\delta}$, i.\ e., $X(t)=X(t+T)$; this can be put as
\begin{equation}
I(a) \equiv \int_{t}^{t+T} {\frac{a + b cos(\delta t' + \delta_{0})}
{\sqrt{1+(a + b cos(\delta t' + \delta_{0}))^{2}}}} dt' = 0, 
\label{ecua20}
\end{equation}
where 
\begin{equation}
a = \frac{u(0)}{\sqrt{1-u(0)^{2}}} \mp \frac
{\pi \epsilon \cos(\delta_{0})}{4 \delta}
\end{equation} 
and $b=\pi \epsilon/4 \delta$. Noticing that, first $I(0)=0$; second,
if $a < -|b|$ $I(a) < 0$ and if $a > |b|$ $I(a) > 0$, and third,
$I'(a) > 0$ for all values of $a$, we arrive at the desired 
result that $I(a) = 0$ if and only if the relation (\ref{ecua8}) [$a = 0$]
holds.
 
\section*{Appendix II}
\label{sec-aII}

In \cite{McL} realistic model of a long Josephson junction is proposed, 
given by the following perturbed sG equation 
\begin{equation}
\phi_{tt} - \phi_{xx} + \sin(\phi) =
-\beta \phi_{t} + \alpha \phi_{xxt} - \gamma ,
\label{aecua1}
\end{equation}
where the term $\beta \phi_{t}$ represents the dissipation due tunneling 
of normal electrons across the barrier, $\alpha \phi_{xxt}$ is the dissipation 
caused by flow of normal electrons parallel to the barrier and 
$\gamma$ is a distributed bias current. In this appendix we apply our 
theory to this problem in order to show that our technique to solve 
the equation for the collective velocity can be applied to other problems. 
In addition, as 
the equation above is realized very approximately by actual Josephson 
devices, by studying it we will providing experimental means of checking 
our results as well as predicting behavior which might have an application
of interest in that context. 

To begin with, we note that if $\alpha = 0$ this equation coincide with 
(\ref{ecua1}) letting $\delta = 0$. Therefore, substituting $\delta = 0$ 
and $\gamma = - \epsilon \sin(\delta_{0})$ in Eqs.\ 
(\ref{ecua99},\ref{ecua999},\ref{ecua9}) we obtain
\begin{equation}
u(t) = \frac{d_{0} \exp(-\beta t) + d}
{\sqrt{1+(d_{0} \exp(-\beta t) + d)^{2}}}, 
%\label{ecua19}
\end{equation}
where $ d_{0} = u(0)/\sqrt{1-u(0)^{2}} - d$ and 
$d=\pm \pi \gamma/4 \beta$. When $t$ goes to infinity, we have
$ u(\infty) \to d/\sqrt{1+d^{2}}$. Notice that this 
expression coincides with 
the equilibrium solution $u_{\infty}$, obtained in \cite{McL}. 
 
For $\alpha \ne 0$ the same procedure as above leads us to a 
first-order ordinary differential equation for $u(t)$, which is given by 
\begin{equation}
\frac{du}{dt} = \pm \frac{\pi \gamma}{4} (1-u^{2})^{3/2} - 
\beta u (1-u^{2}) - \frac{1}{3} \alpha u.
\label{ecuau}
\end{equation}
 If we now introduce the same change of variable as in Sec.\ \ref{sec-cc},
$y(t)=u(t)/\sqrt{1-u(t)^{2}}$, Eq.\  (\ref{ecuau}) becomes 
\begin{equation}
\frac{dy}{dt} = \pm \frac{\pi \gamma}{4} -  
(\beta + \frac{\alpha}{3}) y - \frac{1}{3} \alpha y^{3},
\label{ecuay}
\end{equation} 
Integrating (\ref{ecuay}) we find  
\begin{eqnarray}
\frac{1}{p + x_{1}^{2}} 
\Bigg\{
\ln\Bigg(
c_{1} \frac{y-x_{1}}{\sqrt{\left(y+\frac{x_{1}}{2}\right)^{2}+k_{1}^{2}}}
\Bigg) & - & \nonumber \\
- \frac{3 x_{1}}{2 k_{1}} 
\left[\arctan
\left(\frac{y+\frac{x_{1}}{2}}{k_{1}}
\right)
 - c_{2} 
 \right]
 \Bigg\} & = & - \alpha t,
\end{eqnarray}
%\label{ecua19}
where 
\begin{eqnarray}
3 p & = &  \frac{3 \beta}{\alpha} + 1 \\
2 q & = & \mp \frac{3 \pi \gamma}{4 \alpha}\\
D & = & q^{2} + p^{3}, \\
x_{1} & = &  (\sqrt{D} - q)^{1/3} - \frac{p}
{(\sqrt{D} - q)^{1/3}}, \\
k_{1}^{2} & = & 3 p + \frac{3}{4} x_{1}^{2},\\
c_{1} & = &  \frac{\sqrt{(y(0)+\frac{x_{1}^{2}}{2})^{2} + 
k_{1}^{2}}}{y(0)-x_{1}}, \mbox{ and}\\
c_{2} & = &  \arctan\left(\frac{y(0)+\frac{x_{1}}{2}}{k_{1}}\right). 
\end{eqnarray}

For large values of $t$, $y(\infty) \to x_{1}$, so 
the kink will move with constant velocity $u(\infty) \to 
\frac{x_{1}}{\sqrt{1+x_{1}^{2}}}$, which depends on the parameters $\alpha$, 
$\beta$ and $\gamma$. For example, if $\gamma = 0$ ($q = 0$), 
$\displaystyle u(t)|_{t\to\infty} = 0$, i.e. by the 
influence of both dissipation terms 
($- \beta \phi_{t}$ and $\alpha \phi_{xxt}$) the kink will be stopped.


\begin{thebibliography}{99}
\bibitem{nota} Most important in contexts where soliton stability is a 
crucial factor, as it happens in many applications (Josephson junctions, 
fiber optics). Dynamical solitons are easy much more sensitive to 
perturbations; an archetypal example are those of the Korteweg-de 
Vries equation (see, e.g, \cite{km} and references therein).
\bibitem{not} For $\Omega_0=0.9$, the last point in Fig.\ \ref{graf4}, 
the breather does decay into radiation instead of a into kink-antikink 
pair. We have included this point in the plot because in this case
we could make sure that the breather had indeed disappeared.
\end{thebibliography}
\end{document}